\documentclass[aps,amsmath,amssymb,12pt]{revtex4}
\usepackage{graphicx}
\usepackage{bm}
\usepackage{tcolorbox}

\def\journal#1#2#3#4{{#1} {\bf #2}, #3 (#4)}
\newcommand{\be}{\begin{equation}}
\newcommand{\ee}{\end{equation}}
\newcommand{\bea}{\begin{eqnarray}}
\newcommand{\eea}{\end{eqnarray}}
\newcommand{\hf}{\frac12}
\newcommand{\nn}{\nonumber\\}
\def\eq#1{(\ref{#1})}
\def\la{\langle}
\def\ra{\rangle}

\def\Tr{{\mathrm{Tr}}}
\def\ord#1{{\cal O}\left(#1\right)}
\def\mr#1{{\mathrm{#1}}}
\def\v#1{{\bm{#1}}}

\def\fdd#1#2#3{\frac{\delta^2#1}{\delta#2\delta#3}}
\def\dk{{\Delta k}}

\def\sign{\mr{sign}}

\def\re{\mr{Re}}
\def\im{\mr{Im}}

\begin{document}
\title{Renormalization in Minkowski space-time}

\author{I. Steib$^1$, S. Nagy$^1$, J. Polonyi$^2$}
\affiliation{$^1$Department of Theoretical Physics, University of Debrecen,
P.O. Box 5, H-4010 Debrecen, Hungary}
\affiliation{$^2$Strasbourg University, CNRS-IPHC, \\23 rue du Loess, BP28 67037 Strasbourg Cedex 2, France}
\date{\today}

\begin{abstract}
The multiplicative and the functional renormalization group methods are applied for the four dimensional scalar theory in Minkowski space-time. It is argued that the appropriate choice of the subtraction point is more important in Minkowski than in Euclidean space-time. The parameters of the cutoff theory, defined by a subtraction point in the quasi-particle domain, are complex due to the mass-shell contributions and the renormalization group flow becomes much more involved than its Euclidean counterpart. 
\end{abstract}

\maketitle

\section{Introduction}
Quantum Field Theories are defined in the Minkowski space-time \cite{bogoliubov,bjorken,itzykson,peskin}, and their renormalization, namely the removal of their UV divergences, has been developed accordingly \cite{bogoliubovr,hepp,zimmermann}. The availability of simpler regulators in Euclidean space-time \cite{thooft} and the similarity of the introduction of the  renormalized parameters in Quantum Field Theory with critical phenomena \cite{wilson,amit} led to the recasting of the renormalization group method in imaginary time and developing it further in that context. For instance, most of the functional formalism of the renormalization group method to find non-perturbative solutions of Quantum Field Theory models, are presented in Euclidean space-time \cite{wegner,nicoll,polch,wetterich,ellwanger,morris,berges}. 

Notwithstanding there are several works about functional renormalization group method aiming at the real time dynamics, the first publications being within the framework of the Closed Time Path formalism about the coarse graining \cite{lombardo,dalvit,anastopoulos}, followed by works addressing a quantum dot \cite{gezzi}, open electronic systems \cite{mitra}, transport processes \cite{jacobs}, damping \cite{zanellad}, inflation \cite{zanellai}, quantum cosmology \cite{calzettac}, critical dynamics \cite{canet,mesterhazy,sieberer} and spectral function \cite{huelsmann}. The renormalization group scheme has bee extended to stochastic field theory \cite{zanellac}, too. The 2PI formalism can be used to discover nonthermal fixed points 2PI \cite{bergesfp} and the renormalization group scheme can be transformed to trace the time dependence \cite{gasenzer}. The field theoretical method has been applied for simple generic quantum mechanical problems, such as a system of coupled harmonic oscillators \cite{aoki,kovacs}. The one-loop renormalizability of the scalar model has been worked out on the one-loop level by the help of the more traditional multiplicative renormalization group method \cite{avinash}. There is lately more activity about the real time dynamics within the traditional formalism of quantum field theory by performing the Wick rotation first on the functional renormalization group flow in quantum gravity \cite{manrique}, in QCD \cite{strodthoff}, in the quark-meson model \cite{kamikado1,tripolt1} and on the mesonic spectral functions \cite{kamikado2,strodthoffsp,wambach}. The analytic continuation of the evolution equations was constructed by the help of higher order derivatives \cite{floerchinger} and the analytic continuation was performed on the propagator \cite{pawlowskiacp}. The analytic continuation is avoided in this work and different renormalization group schemes are worked out directly in Minkowski space-time for the four dimensional $\phi^4$ scalar model. 

The main difference between the blocking in imaginary and real time is due to unitarity of the time evolution which makes the Green functions complex. Therefore the running coupling constants, given in terms of the vertex functions evaluated at a subtraction point, become complex, too. It is argued in section \ref{schemess} that the dependence of the complex running parameters on the choice of the subtraction point is stronger for real time dynamics. The results, presented below correspond to subtraction points within the kinematical regime of the quasi-particles. The quasi-particle poles make the Wick rotation singular and the results inaccessible from Euclidean space-time by Wick rotation.

Different multiplicative renormalization group schemes are introduced in section \ref{mults}. The functional renormalization group equation is derived and solved numerically for the bare action and a new crossover scale is identified in section \ref{funcrg}. This work opens more questions than answers, some of them are listed in section \ref{concl}.

\section{Renormalization schemes}\label{schemess}
The renormalization group is designed to find the resolution dependence of the physical quantities. For that end one may follow the evolution of the bare or the effective action action as the function of the maximal resolution, the UV cutoff or the scale where the effective parameters are defined, respectively. The functional renormalization group    scheme is designed to follow the dependence of the parameters of the action as the resolution is changed while keeping the physical content of the theory unchanged. This process requires the adjustment of infinitely many parameters which can in principle be realized in a functional setting of the evolution equation, describing the change of the parameters with the resolution. However the limitation of our analytical possibilities forces us to restrict the cutoff-dependence into a smaller ansatz space of action functionals. This is the point where the efficiency of field theory becomes evident: We face an overdetermined problem, namely to find a large number of physical quantities by the help of the restricted set of parameters of the action, to be solved in a approximate manner by suitable chosen parameters. This choice is made by evaluating the action for an appropriately chosen family of field configurations, called subtraction point.

\subsection{Multiplicative and functional schemes}
The renormalization group method emerged first in high energy phyics in the context of renormalizable theories where the momentum scale of the cutoff, $\Lambda$, is far higher than that of physical observables. The overdetermined problem is solved in the corresponding multiplicative renormalization group schemes by neglecting the irrelevant (non-renormalizable) parameters of the bare, cutoff action. Such a simplification allows us to keep the physics fixed in the limit $\Lambda\to\infty$. The $\Lambda$-independent content of the renormalized theory is expressed in terms of the renormalized parameters, defined by the renormalization conditions, imposed on one-particle irreducible vertex functions evaluated at some suitable chosen external momentum, $p_s$, called subtraction point. Hence the multiplicative renormalization group schemes contain two scales, $\Lambda$ and $p_s$ and either of them can be changed giving rise to renormalized trajectories displaying the $\Lambda$ dependence of the bare theory or the $p_s$ dependence of the renormalized parameters. The multiplicative schemes are usually worked out by the help of the perturbation expansion which can be optimized at different scales. The use of the bare or the renormalized perturbation expansion should be used to recover the $\Lambda$ and the $p_s$ dependence, respectively.

The functional renormalization group method for the bare action has three ingredients. (i) The renormalization condition is replaced by the decrease of the cutoff in the generator functional for the Green function, traditionally called blocking. It consists of the elimination of the field components between the higher and the lower cutoff by using perturbation or loop expansion around the free theory. A decisive advantage of the renormalization group method is the possibility of performing an infinitesimal change of the gliding cutoff and using this change as a small parameter to suppress the higher order contributions. The result is an expansion around the evolving, interactive theory. (ii) The blocking preserves the full dynamics below the new cutoff and gives an extremely complicated functional differential equation leading to non-polynomial, non-local blocked action. This problem is well beyond our capabilities hence one projects the evolution equation into a restricted ansatz space for the blocked action. (iii) Finally the choice of the subtraction point, the way the parameters of the action are extracted from the evolution equation, transforms the functional differential equation for the action into a set of coupled differential equations for the functions or the parameters of the restricted action.

The multiplicative schemes differ from the functional scheme in having two well separated scales, $p_s\ll\Lambda$. As a result two restrictions follow for the former: First, the $\ord{\Lambda^{-n}}$ contributions with $n>0$ are neglected in imposing the renormalization conditions and these schemes are applicable only to renormalizable theories. Second, the renormalized coupling strength $g$ has to be small enough to treat the UV divergences of the form $g\ln\Lambda$ and $g\Lambda^n$ as perturbation. Hence the functional scheme is more suitable to handle competitive scales and crossovers without requiring renormalizability and its limitation in strongly coupled theories arises only from the use of the restricted ansatz space for the blocked action.

\subsection{Minkowski space-time}\label{msrss}
The main novelties of the implementation of the renormalization group method in real time are the complexification of the parameters of the theory and the increased sensitivity on the choice of some conventions. The parameters of the action are given in terms of the vertex functions and become complex owing to the unitarity of the time evolution. In particular, the parameters acquire imaginary part when on-shell intermediate states contribute to the vertex function in question according to the optical theorem. Thus we have twice as many real parameters in the action and even the qualitative, topological features of the flow diagram together with the fixed point and the phase structure may be changed. 

\begin{figure}
\includegraphics[scale=.4]{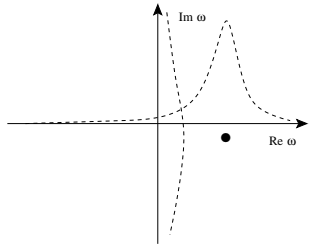}
\caption{A quasi-particle pole, denoted by the heavy dot, on the physical sheet of the complex energy plane of the dressed propagator. The typical dependence of the absolute magnitude of the renormalized propagator on the Euclidean and the Minkowski energy is shown by the dashed lines along the imaginary and real axes, respectively.}\label{quasip}\end{figure}

To find the optimal way to define the running parameters consider for the sake of an example a loop integral of the  renormalized perturbation expansion in a weakly coupled theory. The integral is dominated by the energy range $\omega\sim\re\omega_\v{p}$ ($\omega\sim\im\omega_\v{p}$) in a Minkowski (Euclidean) theory where $\omega_\v{p}=\sqrt{m^2+\v{p}^2}$. The renormalized parameter $m^2$ is defined by the help of the self energy, evaluated at the subtraction point, $\Sigma(p_s)$. The self energy displays a threshold singularity on the mass shell of the quasi-particles, $p^2=m^2$, separating the domain $p^2<m^2$ and $p^2>m^2$ where only virtual and the real particle intermediate states contribute to the self energy, respectively. The subtraction point should be in the latter domain, $p_s^2>m^2$, to pick up the physics of real, propagating particles. The imaginary part of the quasi particle energy is generated by the interactions therefore the quasi-particles have long life-time, $|\im\omega_\v{p}|<|\re\omega_\v{p}|$, c.f. Fig. \ref{quasip}. The upshot is that the quasi-particle peak is narrower in Minkowski space, the increased importance of the choice of the subtraction point.

The usual classification of the running parameters into relevant, marginal and irrelevant classes according to the sign of their scaling exponents is actually incomplete even for Euclidean theories with real coupling because the scaling exponents, the spectrum of the linearized evolution around the fixed point, can be complex and generate spiralling trajectories around the fixed point. It will be seen below that the coefficients of the coupling constant in the evolution equation of a theory in Minkowski space-time are complex and their phase can induce non-monotonic renormalized trajectories and generate crossover scales.

Any massive theory possesses at least one crossover around the mass scale separating the UV and the IR scaling regimes. This is made more abrupt in the symmetry broken phase where a spinodal instability induces a saddle point contribution to the elimination of the UV field components in Euclidean models with real parameters when the quadratic part of the action becomes unstable for the modes to be eliminated, $D^{-1}_p<0$ for $p^2=\Lambda^2$, $D_p$ being the free propagator \cite{tree,vincent}. The complexification of the parameters of the action makes such a saddle point trivial in Minkowski space-time. In fact, the Fresnel integral, used for a free theory is well defined as long as the integrand is bounded, $\im D^{-1}_p>0$, allowing either sign for $\re D^{-1}_p$ with vanishing saddle point. This generalises to higher order terms in the local potential, $g_n=\delta^nS/\delta\phi^n_x$. The singularity of the Euclidean theory with real parameters at $\re g_n=0$ is smoothened out in Minkowski space-time however new characteristic scales $\Lambda_n$ are  generated where $\im g_n=0$ by the appearence of complex saddle points. The path integral \eq{blrel} converges and is well defined as long as the imaginary part of the highest order coupling constant is negative. We find two further crossover scales at $\Lambda_{sp}$ and $\Lambda_L$, the left over of the spinodal transition of the Euclidean theory with real parameters where $\re D^{-1}_p$ reaches its peak for $p^0=0$ and $|\v{p}|=\Lambda_{ps}$ and the remains of the Landau pole, a peak of $|g_4|$, respectively. 

One could in principle extend the Euclidean theories for complex parameters. Such theories display crossovers at $\re g_n=0$ and are well defined as long as the real part of the highest order coupling constant is positive. This possibility is not pursued in this work because the quasi-particle dynamics is captured better in Minkowski space-time.

\section{Multiplicative renormalization group}\label{mults}
The Lagrangian of the scalar model, defined by the renormalized perturbation expansion is
\be
L=\hf\partial_\mu\phi\partial^\mu\phi-\frac{m^2}2\phi^2-\frac{g}{4!}\phi^4+\frac{\delta Z}2\partial_\mu\phi\partial^\mu\phi-\frac{\delta m^2}2\phi^2-\frac{\delta g}{4!}\phi^4
\ee
where $m^2$ and $g$ are the renormalized parameters and the counterterms, parametrized by $\delta Z=\ord{g^2}$, $\delta m^2=\ord{g}$ and $\delta g=\ord{g^2}$ are to remove the UV divergences from the renormalized, physical quantities. The Lagrangian of the bare theory is
\be
L=\hf\partial_\mu\phi_B\partial^\mu\phi_B-\frac{m_B^2}2\phi_B^2-\frac{g_B}{4!}\phi_B^4
\ee 
with $Z=1+\delta Z$, $Zm_B^2=m^2+\delta m^2$ and $Z^2g_B=g+\delta g^2$ where $\phi_B=\sqrt{Z}\phi$. The usual $\epsilon$ prescription is realized by requiring $\im m_B^2,\im m^2<0$. The regulator is sharp momentum cutoff, $|\v{p}|<\Lambda$ and the status of the Lorentz symmetry is not pursued in this work.

\subsection{Renormalization conditions}
The renormalization conditions connect the bare and the renormalized theories. The renormalized mass and the wavefunction renormalization constants are defined by the help of the exact propagator \cite{georgi},
\be
\la T[\phi_{-p}\phi_p]\ra=\frac{i}{p^2-m_B^2-\Sigma_B(p^2)},
\ee
evaluated around the subtraction momentum $p^\mu\approx p^\mu_s=(\omega_s,\v{0})$,
\be
p^2-m_B^2-\Sigma_B-p^2\partial_{p^2}\Sigma_B=\frac{p^2-m^2}Z,
\ee
with $\Sigma_B=\Sigma_B(p^2_s)$, yielding
\be
m^2=Z(m_B^2+\Sigma),
\ee
and
\be
Z=\frac1{1-\partial_{p^2}\Sigma}.
\ee
The renormalized coupling constant is defined by the help of the four point vertex function,
\be
Z^2g=\Gamma_4(p_s,p_s,p_s,p_s).
\ee

\begin{figure}
\includegraphics[scale=.2]{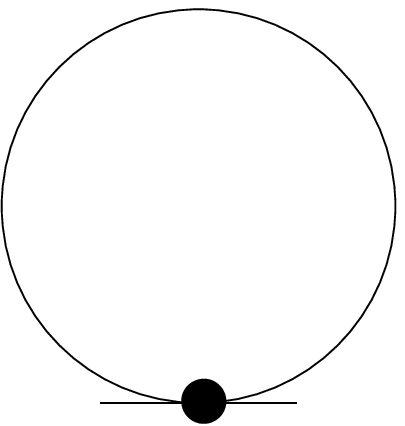}\hskip1cm
\includegraphics[scale=.2]{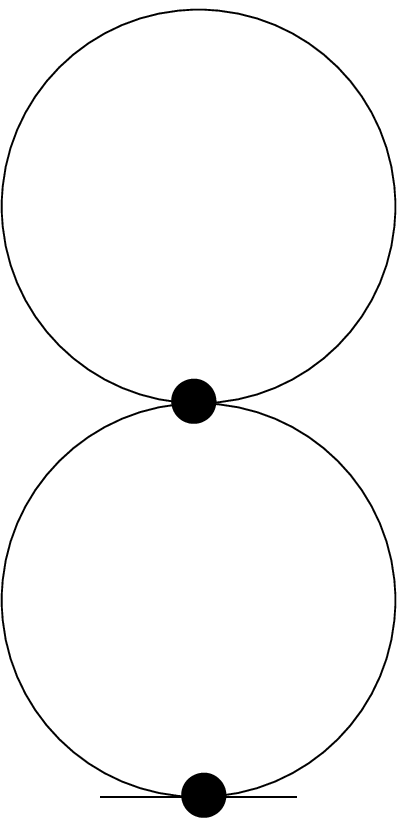}\hskip1cm
\includegraphics[scale=.2]{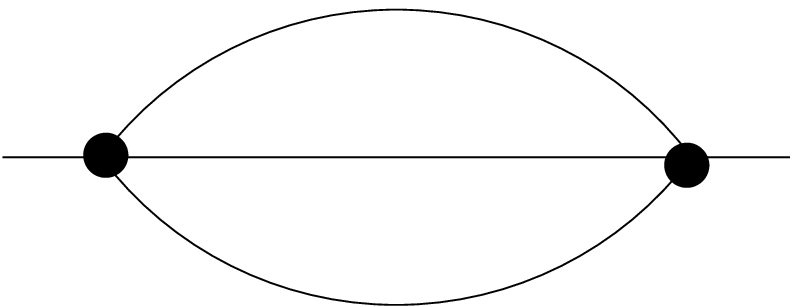}
\caption{Graphs of the two-loop self energy.}\label{g2graphs}\end{figure}

\begin{figure}
\includegraphics[scale=.2]{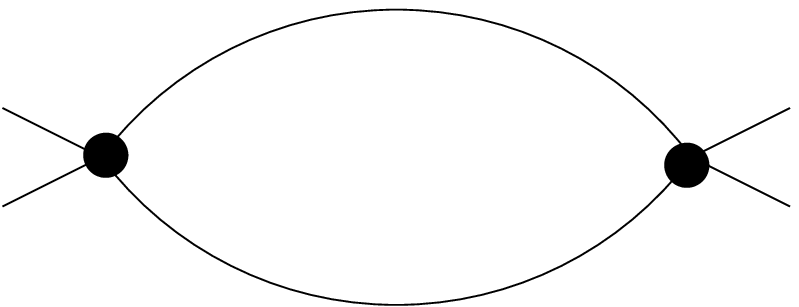}\hskip1cm
\includegraphics[scale=.2]{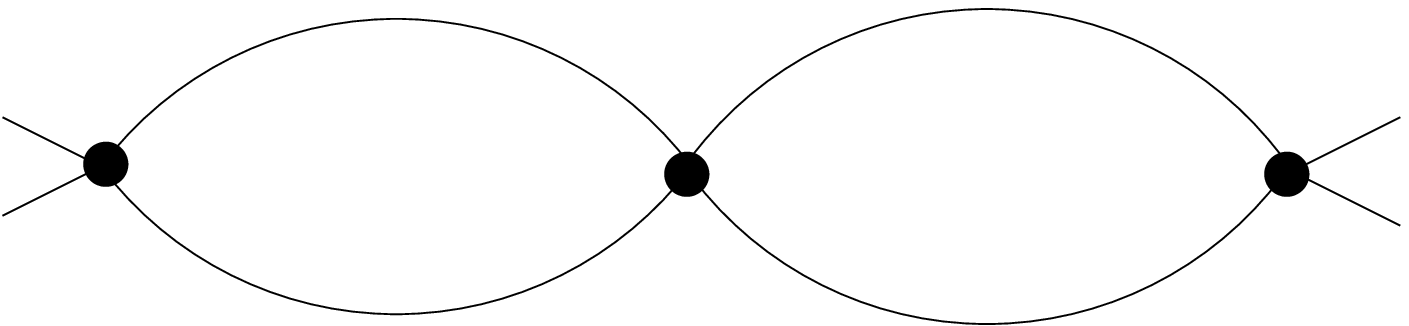}\hskip1cm
\includegraphics[scale=.2]{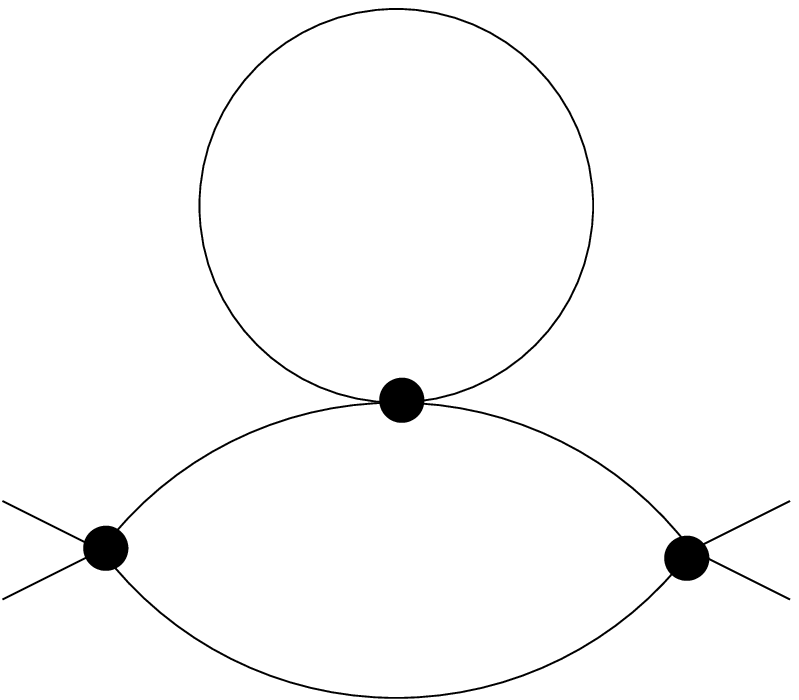}\hskip1cm
\includegraphics[scale=.3]{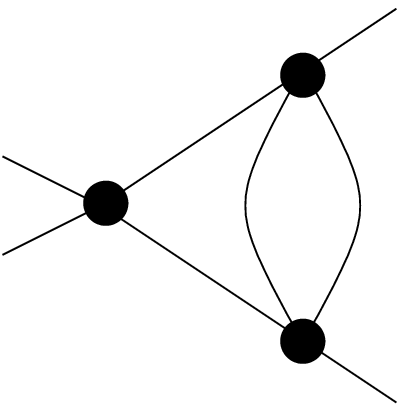}
\caption{Graphs of the fourth order vertex function.}\label{g4graphs}\end{figure}

The two-loop expressions for the second and the fourth order vertex functions receive contributions from the graphs, depicted in Figs. \ref{g2graphs} and \ref{g4graphs}, respectively. The resulting renormalization conditions are
\bea\label{rencond}
m^2&=&\frac{m_B^2+\frac{i}2L^{(2)}_{1B}g_B-(\frac14L^{(2)}_{1B}L^{(2)}_{2B}+\frac16L^{(2)}_{3B})g_B^2}{1-\partial_{p^2}\Sigma_B}+\ord{g_B^3},\nn
g&=&g_B+\frac{i}2L^{(4)}_1g_B^2+\left(\frac13\partial_{\omega_s^2}L^{(2)}_{3B}-\hf L^{(4)}_{2B}L^{(2)}_{1B}-\hf L^{(4)}_{3B}-\frac14L^{(4)2}_{1B}\right)g_B^3+\ord{g_B^4},\nn
\frac1Z&=&1+\frac{g_B^2}6\partial_{\omega_s^2}L^{(2)}_{3B}+\ord{g_B^3},
\eea
where
\bea\label{loopints}
L^{(2)}_{1B}&=&\int\frac{d^4q}{(2\pi)^4}\frac1{q^2-m_B^2},\nn
L^{(2)}_{2B}&=&\int\frac{d^4q}{(2\pi)^4}\frac1{(q^2-m_B^2)^2},\nn
L^{(2)}_{3B}&=&\int\frac{d^4q_1d^4q_2}{(2\pi)^8}\frac1{(q_1^2-m_B^2)(q_2^2-m_B^2)[(p_s-q_1-q_2)^2-m_B^2]},\nn
L^{(4)}_{1B}&=&3\int\frac{d^4q}{(2\pi)^4}\frac1{(q^2-m_B^2)[(p_s-q)^2-m_B^2]},\nn
L^{(4)}_{2B}&=&3\int\frac{d^4q}{(2\pi)^4}\frac1{(q^2-m_B^2)^2[(p_s-q)^2-m_B^2]},\nn
L^{(4)}_{3B}&=&6\int\frac{d^4q_1d^4q_2}{(2\pi)^8}\frac1{(q_1^2-m_B^2)(q_2^2-m_B^2)[(2p_s-q_1)^2-m_B^2][(p_s-q_1-q_2)^2-m_B^2]}.
\eea
The integration over the energy is for $-\infty<q^0<\infty$ and the spatial momentum is restricted to $|\v{q}|<\Lambda$.

It is sometime advantageous to use the renormalized perturbation expansion. For this end we write the bare propagator
\be
\frac1{q^2-m^2_B}=D_{Rq}[1+D_{Rq}(\Sigma_B+q^2\partial_{p^2}\Sigma_B)+\ord{g^2}],
\ee
in terms of the renormalized propagator $D_{Rq}=1/(q^2-m^2)$ in the loop integrals. The corrections represent the counterterms for the mass and the wavefunction renormalization constant. The renormalized loop integrals $L^{(j)}_{kR}$ are obtained from $L^{(j)}_{kB}$ by the replacement $m_B^2\to m^2$ in the two-loop calculation except $L^{(2)}_{1B}\approx L^{(2)}_{1R}+g\ell^{(2)}_1$ and $L^{(4)}_{1B}\approx L^{(4)}_{1R}+g\ell^{(4)}_1$ where the $\ord{g}$ counterterms appear,
\bea
g\ell^{(2)}_1&=&\int_qD^2_{Rq}(\Sigma_B+q^2\partial_{p^2}\Sigma_B),\nn
g\ell^{(4)}_1&=&3\int_qD_{Rq}D_{Rp_s-q}[D_{Rq}(\Sigma_B+q^2\partial_{p^2}\Sigma_B)+D_{Rp_s-q}(\Sigma_B+(p_s-q)^2\partial_{p^2}\Sigma_B)].
\eea

It is worthwhile to distinguish the historical definition of renormalizability from the more careful one. The renormalizability was first meant as the possibility of choosing the counterterms in such a cutoff-dependent manner that the UV divergences cancel in the vertex functions in each order of the perturbation expansion and is expressed by using power counting for the primitive non-overlapping divergences. However this is not enough to perform the limit $\Lambda\to\infty$. In fact, the renormalization conditions, \eq{rencond}, are non-linear in the bare parameters and the cancellation of the UV divergences does not guarantee the existence their solution. It may happen that the there are no finite bare parameters to satisfy the renormalization conditions beyond a certain value of the cutoff. Such a value of the cutoff is usually called Landau pole.

\subsection{Gliding cutoff scale with renormalized perturbation expansion}\label{cutren}
The renormalization group scheme where the renormalized trajectory can be found analytically follows the evolution of the bare parameters as the cutoff is changed by the help of the renormalized perturbation expansion. The loop integrals of the renormalized perturbation expansion contain the renormalized mass which is kept constant when the cutoff is moved hence the beta functions of this scheme are independent of the bare mass, allowing their integration in closed form.

To find the dependence of the bare parameters on the cutoff for a fixed renormalized theory we first calculate the derivative of the renormalization conditions \eq{rencond} with respect to $t=\ln\Lambda/\Lambda_0$ by keeping the renormalized parameters fixed. The resulting system of equations is then solved for the bare mass-independent beta functions,
\bea
\partial_tm_B^2&=&\beta_{m_B^2}^{(1R)}g_B+\beta_{m_B^2}^{(2R)}g_B^2+\ord{g_B^3},\nn
\partial_tg_B&=&\beta_{g_B}^{(2R)}g_B^2+\beta_{g_B}^{(3R)}g_B^3+\ord{g_B^4},\nn
\partial_tZ&=&\beta_{ZB}^{(2R)}g_B^2+\ord{g_B^3},
\eea
where
\bea
\beta_{m_B^2}^{(1R)}&=&-\frac{i}2\partial_tL^{(2)}_{1R},\nn 
\beta_{m_B^2}^{(2R)}&=&-\frac{i}2\partial_t\ell^{(2)}_1+\frac14\partial_tL^{(2)}_{1R}L^{(2)}_{2R}+\frac14L^{(2)}_{1R}\partial_tL^{(2)}_{2R}+\frac16\partial_tL^{(2)}_{3R}-\frac14L^{(2)}_1\partial_tL^{(4)}_{1R}+\frac16m^2\partial_t\partial_{\omega_s^2}L^{(2)}_{3R},\nn
\beta_{g_B}^{(2R)}&=&-\frac{i}2\partial_tL^{(4)}_{1R},\nn
\beta_{g_B}^{(3R)}&=&-\frac{i}2\partial_t\ell^{(4)}_1-\frac13\partial_t\partial_{p^2}L^{(2)}_{3R}+\hf\partial_tL^{(4)}_{2R}L^{(2)}_{1B}+\hf L^{(4)}_{2R}\partial_tL^{(2)}_{1R}+\hf\partial_tL^{(4)}_{3R},\nn
\beta_Z^{(2R)}&=&-\frac16\partial_t\partial_{\omega_s^2}L^{(2)}_{3R}.
\eea

The integration of the one-loop beta functions is trivial: The coupling constant follows the trajectory
\be
g_B(\Lambda)=\frac{g_B(\Lambda_0)}{1-g_B(\Lambda_0)\beta^{(2R)}_{g_B}\ln\frac{\Lambda}{\Lambda_0}},
\ee
with $\beta^{(2R)}_{g_B}>0$ and displays a Landau pole at $\Lambda_L=\Lambda_0\exp(1/g_B(\Lambda_0)\beta^{(2R)}_{g_B})$. The mass, given by
\be
m^2_B(\Lambda)=m^2_B(\Lambda_0)-\frac{\beta_{m_B^2}^{(1R)}}{\beta^{(2R)}_{g_B}}\ln\left(1-g_B(\Lambda_0)\beta^{(2R)}_{g_B}\ln\frac{\Lambda}{\Lambda_0}\right),
\ee
$\beta_{m_B^2}^{(1R)}<0$, approaches $-\infty$ at the Landau pole. Finally, the wavefunction renormalization constant is kept constant by the one-loop beta functions, $Z(\Lambda)=Z(\Lambda_0)$. We rely on the renormalized perturbation expansion hence these results are valid only in a finite cutoff range, as long the radiative corrections are small, $g_B(\Lambda_0)\Lambda^2\ll m_B^2(\Lambda_0)$. The two-loop contributions to the beta functions do not change the scaling laws in an important manner within this cutoff range. 

It is pointed out in section \ref{renren} that the coefficients of the $g_B$ powers in the beta functions are actually complex hence the Landau pole of the one-loop beta functions is missed, $|g_B|$ displays a Landau peak at $\Lambda_L=\Lambda_0\exp(1/\re(g_B(\Lambda_0)\beta^{(2R)}_{g_B}))$ and the theory is asymptotically free, assuming the absence of saddle points and the convergence of the path integral.

The price paid for the bare mass-independence of this scheme is the optimisation of the perturbation expansion at the fixed, physical scale. To realize an expansion around the evolving interacting theory we need an optimisation at the gliding scale, to be realized by following either the bare parameters as the function of the cutoff or the subtraction scale dependence of the renormalized parameters using the bare or the renormalized perturbation expansion, respectively.

\subsection{Gliding cutoff scale with bare perturbation expansion}\label{cutbare}
The multiplicative scheme which is the closest to the functional renormalization group method is where the cutoff dependence of the bare parameters are established by the bare perturbation expansion. The starting point is the calculation of the derivative of the renormalization conditions \eq{rencond} with respect to $t=\ln\Lambda/\Lambda_0$ by keeping the renormalized parameters fixed. The resulting system of equations is then solved for the mass-dependent beta functions,
\bea
\partial_tm_B^2&=&\beta_{m_B^2}^{(1B)}g_B+\beta_{m_B^2}^{(2B)}g_B^2+\ord{g_B^3},\nn
\partial_tg_B&=&\beta_{g_B}^{(2B)}g_B^2+\beta_{g_B}^{(3B)}g_B^3+\ord{g_B^4},\nn
\partial_tZ&=&\beta_{ZB}^{(2B)}g_B^2+\ord{g_B^3},
\eea
where
\bea\label{lblist}
\beta_{m_B^2}^{(1B)}&=&-\frac{i}2\partial_tL^{(2)}_{1B},\nn
\beta_{m_B^2}^{(2B)}&=&-\frac14\partial^2_{m_B^2}L^{(2)}_{1B}\partial_tL^{(2)}_{1B}+\frac14\partial_tL^{(2)}_{1B}L^{(2)}_{2B}+\frac14L^{(2)}_{1B}\partial_tL^{(2)}_{2B}+\frac16\partial_tL^{(2)}_{3B}-\frac14L^{(2)}_1\partial_tL^{(4)}_{1B}\nn
&&+\frac16m_B^2\partial_t\partial_{\omega_s^2}L^{(2)}_{3B},\nn
\beta_{g_B}^{(2B)}&=&-\frac{i}2\partial_tL^{(4)}_{1B},\nn
\beta_{g_B}^{(3B)}&=&-\frac13\partial_t\partial_{p^2}L^{(2)}_{3B}-\frac14\partial_tL^{(2)}_{1B}\partial_{m_B^2}L^{(4)}_{1B}+\hf\partial_tL^{(4)}_{2B}L^{(2)}_{1B}+\hf L^{(4)}_{2B}\partial_tL^{(2)}_{1B}+\hf\partial_tL^{(4)}_{3B},\nn
\beta_{ZB}^{(2B)}&=&-\frac16\partial_t\partial_{\omega_s^2}L^{(2)}_{3B}.
\eea

Let us discuss separately the case of real and complex parameters. Real parameters in Minkowski space-time mean $\im g_B=0$, $\im m_B^2=-\epsilon$ and the one-loop contributions to the beta functions are real as $\epsilon\to0$. In fact, a loop integral with $\im m_B^2=-\epsilon$ may develop finite imaginary part only if the integration contour passes close to the mass-shell. However $\partial_\Lambda$ removes the momentum integration and we are left with the real integrand,
\bea
\beta_{m_B^2}^{(1)}&=&-\frac1{8\pi^2}\Lambda\partial_\Lambda\int_0^\Lambda dq\frac{q^2}{\omega_{Bq}}\nn
&=&-\frac1{8\pi^2}\frac{\Lambda^3}{\omega_{B\Lambda}},\nn
\beta_{g_B}^{(2)}&=&\frac3{16\pi^2}\Lambda\partial_\Lambda\int_0^\Lambda\frac{dqq^2}{\omega_{Bq}}\frac1{q^2+m_B^2-\frac{\omega^2_s}4}\nn
&=&\frac3{16\pi^2}\frac{\Lambda^3}{\omega_{B\Lambda}}\frac1{\Lambda^2+m_B^2-\frac{\omega^2_s}4},
\eea
where $\omega_{Bq}=\sqrt{m_B^2+q^2}$. This result holds only for $O(d-1)$ invariant momentum cutoff since $\partial_\Lambda$ does not completely eliminate the momentum integration for non rotational invariant cutoff, e.g. $|p_j|<k$, $j=1,\ldots,d-1$ if $d-1\ge2$. There is an UV Landau pole and the one-loop beta functions can be integrated for $\Lambda<\Lambda_L$. 

There is a qualitative difference between the one- and the two-loop level results in real time owing to the mass-shell singularity. The point is that $\partial_\Lambda$ does not eliminate the momentum integration in the higher loop integrals even in the case of the $O(d-1)$ invariant cutoff. In particular it is easy to see that $\partial_\Lambda L^{(2)}_{3B}$ and $\partial_\Lambda L^{(4)}_{3B}$ contain momentum integration which generates finite imaginary part for the beta functions and the trajectory becomes complex even for real initial value of $m_B^2(\Lambda_0)$ and $g_B(\Lambda_0)$ when $\omega^2_s>4m_B^2$. The renormalized trajectory remains real and the analytic extension of the Euclidean model when the subtraction point is placed in the virtual regime, $\omega^2_s<4m_B^2$. The mass-shell singularities, indicating the presence of real quasi-particle excitations, render the Wick rotation non-analytic for the physically better justified placing of the subtraction point in the quasi-particle domain. The internal line mass-shell contributions to the beta functions make the renormalized trajectory complex even if we start with real initial conditions.

\subsection{Gliding subtraction scale}\label{renren}
The traditional use of the multiplicative scheme produces the subtraction scale dependence of the renormalized parameters for a fixed cutoff theory by the help of the renormalized perturbation expansion. The solution of the derivative of the renormalization conditions \eq{rencond} with respect to $t=\ln p_s/p_{s0}$ for the derivative of the renormalized parameters and $Z$ is
\bea
\partial_tm^2&=&\beta_{m^2}^{(2)}g^2+\ord{g^3},\nn
\partial_tg&=&\beta_g^{(2)}g^2+\beta_g^{(3)}g^3+\ord{g^4},\nn
\partial_tZ&=&\beta_Z^{(2)}g^2+\ord{g^3},
\eea
where
\bea
\beta_{m^2}^{(2)}&=&-\frac14\partial_tL^{(2)}_{1R}\partial_{m^2}L^{(2)}_{1R}+\frac{i}2\partial_t\ell^{(2)}_1-\frac14\partial_tL^{(2)}_{1R}L^{(2)}_{2R}-\frac14L^{(2)}_{1R}\partial_tL^{(2)}_{2R}-\frac16\partial_tL^{(2)}_{3R}+\frac14\partial_tL^{(2)}_{1R}L^{(4)}_{1R}\nn
&&-\frac13m^2\partial_t\partial_{\omega_s^2}L^{(2)}_{3R},\nn
\beta_g^{(2)}&=&\frac{i}2\partial_tL^{(4)}_{1R},\nn
\beta_g^{(3)}&=&\frac13\partial_{\omega_s^2}\partial_tL^{(2)}_{3R}-\frac14\partial_tL^{(2)}_{1R}\partial_{m^2}L^{(4)}_{1R}+\frac{i}2\partial_t\ell^{(4)}_1-\hf\partial_tL^{(2)}_{1R}L^{(4)}_{2R}-\hf L^{(2)}_{1R}\partial_tL^{(4)}_{2R}-\hf\partial_tL^{(4)}_{3R},\nn
\beta_Z^{(2)}&=&-\frac16\partial_t\partial_{\omega_s^2}L^{(2)}_{3R}.
\eea
The $\ord{g}$ term is missing form the beta function of the renormalized mass due to the subtraction scale independence of the leading order graph in the mass renormalization condition. The $p_s$-dependent loop integrals are complex, e.g. 
\be
\beta_g^{(2)}=i\frac32\int\frac{dq}{(2\pi)^4}\frac{\omega^2_s(\omega_s-q^0)}{(q^2-m^2)[(p_s-q)^2-m^2]^2}
\ee
develops imaginary part for $\omega_s>2m$ and $\im\partial_t\partial_{\omega_s^2}L^{(2)}_{3R}\ne0$ when $\omega_s>3m$ and make $m^2$ and Z complex, respectively. To pick up the quasi-particle dynamics we need a subtraction scale $\omega_s$ sightly above $3m$. The resulting complex $m^2$ reflects the finite life-time of quasi-particles and gives the argument about the complex nature of the beta function coefficients of section \ref{cutren}.

\section{Functional renormalization group}\label{funcrg}
A more detailed description of the functional renormalization group method in Minkowski space-time is given below by introducing the renormalization group scheme, indicating the calculation of the evolution equation of the bare action and presenting some typical renormalized trajectory.

\subsection{Renormalization group scheme}
The main ingredients of this scheme, mentioned in section \ref{schemess} are the following:

(i) {\em Blocking:} The blocking is defined without any reference to an UV fixed point by decreases the cutoff, $k\to k-\dk$, the step size, $\dk$, playing the role of a small parameter. The field variable,
\be
\phi_p=\int d^dxe^{ipx}\phi_x,
\ee
in the Fourier space is split into the sum $\phi\to\phi+\varphi$, where $\phi$ ($\varphi$) belongs to the IR, under the cutoff (UV, beyond the cutoff). The cutoff consists of a choice of the support $P_k$ of the IR field in the momentum space. The change of the action $S_k[\phi]$, corresponding to the cutoff $k$, is given by the blocking relation,
\be\label{blrel}
e^{iS_{k-\dk}(\phi)}=\int D[\varphi]e^{iS_k[\phi+\varphi]}.
\ee
The integration over the UV field can be carried out within the framework of the loop expansion. The leading order expression is
\be
e^{iS_{k-\dk}(\phi)}=\int D[\varphi]e^{iS_k[\phi+\varphi_0]+\frac{i}2\varphi\fdd{S_k[\phi+\varphi_0]}{\varphi}{\varphi}\varphi+\ord{\dk}},
\ee
where $\varphi_0$ stands for the saddle point and the higher order contributions are suppressed by the small parameter, the $\ord{\dk}$ volume of the integration domain in the momentum space. We assume $\varphi_0=0$ below for the sake of simplicity and find the exact one-loop evolution equations
\be\label{eveq}
\dot S[\phi]=-i\frac{k}2\Tr\ln\left[\fdd{S}{\phi}{\phi}\right],
\ee
where $\dot f=\partial_\tau f$, $\tau=\ln(k/k_{in})$, $k_{in}$ being the initial value of the cutoff and the trace is over the UV field space. The initial conditions for the differential equation \eq{eveq} are imposed at $k=k_{in}$. Note that the complex conjugate of the action follows an inverted evolution in real time, more precisely $S[\phi]$ and $-S^*[\phi]$ obey the same evolution equation, reminiscent of the time inversion of the equation of motion.

(ii) {\em Restricted ansatz space:} We truncate the functional differential equation \eq{eveq} onto the $\ord{\partial^2}$ level of the gradient expansion,
\be\label{blact}
S[\phi]=\int d^dx\left[\frac{Z_t(\phi_x)}2(\partial_0\phi_x)^2-\frac{Z_s(\phi_x)}2(\v{\partial}\phi_x)^2-U(\phi_x)\right],
\ee
and the local functions
\bea\label{locfuncts}
Z_t(\phi)&=&\sum_{n=0}^{N_Z}\frac{z_{t,2n}}{2n!}\phi^{2n},\nn
Z_s(\phi)&=&\sum_{n=0}^{N_Z}\frac{z_{s,2n}}{2n!}\phi^{2n},\nn
U(\phi)&=&\sum_{n=0}^{N_U}\frac{g_{2n}}{2n!}\phi^{2n},
\eea
$z_{t,n}=z_{t,n,r}+iz_{t,n,i}$, $z_{s,n}=z_{s,n,r}+iz_{s,n,i}$, $g_n=g_{n,r}+ig_{n,i}$, are restricted to $N_Z=1$ and $N_U=2$ in this work.

(iii) {\em Subtraction point:} The parameters of the blocked action are defined by evaluating the evolution equation at a subtraction point, the IR field $\phi_{s,x}=\Phi+\chi_x$, being the sum of a homogeneous and an inhomogeneous  components, $\Phi$ and $\chi_p=\Psi(2\pi)^d[\delta(p-p_s)+\delta(p+p_s)]/2$, $p^\mu_s=(\omega_s,\v{0})$, respectively. The sum over plane waves with opposite momentum reflects the impossibility of distinguishing particles and anti-particles in the absence of conserved charge. Note that these renormalization conditions are different than those, used for the multiplicative schemes. In fact, while eqs. \eq{rencond} involve the vertex functions evaluated at the subtraction scale, $p_s$, the blocked action, evaluated at $\phi_s$, defines the parameters by the sum of the vertex functions at external moments $p_s$ and $-p_s$. The simpler renormalization conditions, build on $p_s$ alone, are available only for complex fields where the particle and anti-particles can be distinguished.

Few remarks are in order about the scheme: 

(a) One should retain the possible non-trivial saddle points of the blocking \eq{blrel}. Being space-time dependent  they induce non-local contributions to the blocked action. Hence the use of the gradient expansion implies the omission of the non-trivial saddle points. 

(b) The representation of the local potential by a polynomial of finite order can only be justified for weakly coupled theories. 

(c) We need $g_{N_U,r}<0$ to assure the convergence of the path integral. The boundedness of the energy from below requires $g_{N_U,r}\ge0$ in a theory with real parameters. The issue of stability becomes more involved with complex parameters because the quasi-particles of finite life-time can not destabilize the system with unbounded energy and one has to take into account the radiation energy loss to the environment to construct the asymptotic states. This problem is postponed to a later time and the convergence of the path integral, $g_{N_U,i}<0$, is used in this calculation as the only restriction on the parameters. 

(d) While the loop expansion requires $g_{2,i}\ne0$ in Minkowski space-time the evolution equation remains integrable in the cutoff at $g_{2,i}=0$.

(e) The IR field is the same on both sides of the evolution equation, \eq{blrel}, hence its cutoff-dependence does not contribute to the left hand side of eqs. \eq{eveq}. 

(f) The evolution equation of the functional scheme contains only one-loop contributions hence the closest we can go to the leading order multiplicative scheme of section \ref{cutbare} is to restrict the blocked action to a renormalizable one. The impact of the higher loop contributions to the multiplicative scheme beta functions on the renormalized trajectory can partially be recovered in the functional scheme by allowing non-renormalizable terms in the blocked action. For instance the contributions to the last graphs of Figs. \ref{g2graphs} can be found by considering following two successive blocking steps, $k\to k-\dk\to k-2\dk$, and identifying the contribution to the evolution of $g_4$ and $g_2$ in the first and the second step, respectively. But the functional and the multiplicative schemes remain different owing to the keeping of the full cutoff dependence in the former and the ignoring to the $\ord{k^{-n}}$, $n>0$ contributions in the renormalization condition of the latter.

(g) The return to real time in the renormalization group equation poses an unexpected problem, the difficulty of maintaining the boost invariance in non-perturbative schemes \cite{boostinv}. This issue is circumvented here by relying non-relativistic cutoff, $P_k=\{(p^0,\v{p})||p^0|<\Omega_k,|\v{p}|<k\}$, $\Omega_k$ being a suitably chosen non-decreasing function of $k$ and leave the issue of a possible restoration of the boost symmetry in the renormalized theory for a later time. To minimize the symmetry breaking effects of the  cutoff the running action is projected back to the symmetric form \eq{blact} after the blocking. 

(h) One tends to use the simpler procedure by avoiding the cutoff in energy, $\Omega_k=\infty$ but the energy integrals have to regulated if the wavefunction renormalization constant is field-dependent. This can be understood by performing the Fourier transform in space, $\phi_x\to\phi_{t,\v{p}}$, and treating $\phi_{t,\v{p}}$ as the coordinate of a quantum mechanical problem. The usual quantum mechanical Hamiltonian without operator mixing, i.e. of the form $H=p^2/2m+U(x)$, generate UV finite dynamics. However the field-dependent wave function renormalization constant appears as operator mixing, the emergence of the product of momentum $p$ and coordinate $x$ in the Hamiltonian and leads to UV divergent integrals in perturbation expansion \cite{rgqm}.

(i) The energy cutoff introduces non-causal effects at the cutoff scale. But this is a known feature of any regularization procedure where the cutoff reflects our ignorance and the dynamics is reproduced only in an approximative manner at the cutoff scale.

(j) The loop-integral in the evolution equation is over the set $\Delta P_k=P_k\setminus P_{k-\dk}$ of the momentum space and the evolution of $Z$ comes from the overlap of two such a region, shifted towards each other. To allow a finite volume for $\Delta P_k$ have need a cutoff where $P_k$ has a curvature-free hypersurface. This is realized by our non-relativistic regulator.

\subsection{Evolution equation}
The left hand side of the evolution equation becomes
\be
L=V^{(d)}\left(\frac{Z}4\Psi^2\omega_s^2-\frac{Z''}{32}\Psi^4\omega_s^2-U-\frac{U''}4\Psi^2-\frac{3U^{(4)}}{192}\Psi^4\right)+\left(\frac{Z}2-\frac{Z''}8\Psi^2\right)\int d^dx(\partial\chi_x)^2
\ee 
in the $\ord{\chi^2}$ approximation where $V^{(d)}$ denotes the space-time volume and the coefficients of the series \eq{locfuncts} are defined by expanding around $\Phi$. The second functional derivative of the action,
\be
\fdd{S}{\chi_p}{\chi_q}=D^{-1}_p\delta_{p,q}-\Sigma_{p,q}
\ee
with $D^{-1}_p=Zp^2+U''$, $\Sigma=\Sigma^{(1)}+\Sigma^{(2)}$,
\bea\label{selfe12}
\Sigma^{(1)}_{p,q}&=&-\int d^dr\delta(p-q+r)\chi_r[Z'(r^2+pq)+U^{(3)}],\nn
\Sigma^{(2)}_{p,q}&=&-\hf\int\frac{d^dr}{(2\pi)^d}\chi_r\chi_{p-q-r}[Z''(r^2+r(p-q)+pq)+U^{(4)}],
\eea
with $U^{(n)}=\partial^n_\Phi U(\Phi)$ gives the right hand side
\bea\label{right}
R&=&-\hf\int\frac{d^dp}{(2\pi)^d}\biggl\{\Theta(\Omega-|p^0|)\delta(k-|\v{p}|)\left[D\Sigma^{(2)}+\frac{(D\Sigma^{(1)})^2}2\right]_{p,p}\nn
&&+\dot\Omega\sum_{\sigma=\pm}\delta(\sigma\Omega-p^0)\Theta(k-|\v{p}|<k)\left[D\Sigma^{(2)}+\frac{(D\Sigma^{(1)})^2}2\right]_{p,p}\biggr\}
\eea
up to $\ord{\chi^2}$. The identification of the coefficients of different powers of $\Psi$, $\chi$ and $\omega_s$ yields a linear equation for the beta functions for the parameters $g_n$, $z_{s,n}$ and $z_{t,n}$, too lengthy to record here. 

However the simplified case $Z=1$ and $\Omega=\infty$ is easy to present,
\bea\label{olbfs}
\dot g_2&=&\sign(g_{2i})\alpha_{d-1}\frac{k^{d-1}}{4\omega_k}g_4,\nn
\dot g_4&=&\sign(g_{2i})\alpha_{d-1}\frac{12k^{d-1}}{\omega_k}g_4^2\left[\frac1{2(\omega_s^2-\omega_k^2)}-\frac1{\omega_k^2}\right]
\eea
where $\alpha_d=2\pi^{d/2}/(2\pi)^d\Gamma(d/2)$ and $\omega_k=\sqrt{z_{s,0}k^2+g_2}$. The potentially singular first term in the beta function of $g_4$ emerges from the second graph of Fig. \ref{g4graphs} in a manner similar to the nesting of the Fermi surfaces, the common origin being a narrow shell of three-momentum integration. The beta functions represent the change of the parameters of the theory during an infinitesimal change of the cutoff, $\Delta g_n=\dk\beta_n$ and the continuity of the renormalized trajectory, $\Delta g_n\to0$ as $\dk\to0$, is a central point of the of the renormalization group method and can be proven rigorously in Euclidean space-time \cite{israel,enter}. It explains for instance that despite the regularity of the  physical laws at any given scale the singularities of a critical system do arise from the diverging scale window between the microscopic cutoff. The singularity at $\omega_s^2=\omega_k^2$ is generated by the mass-shell singularities, a real time effect. The discontinuity at $g_{2,i}=0$ is a real time effect, as well and is related to the ill-defined nature of the Minkowski space-time loop integrals with $g_{2,i}=0$. Nevertheless $g_{2,i}$ may cross zero as long as $g_{4,i}<0$, only the perturbation expansion produces an integrable singularity in the form of a saddle point.

The subtraction point
\be
\omega_s=\frac{c_s}{\sqrt{|z_{t,0,r}|}}\begin{cases}\sqrt{|z_{s,0,r}k^2+g_{2,r}|}&|z_{s,0,r}k^2+g_{2,r}|>1\cr k\sqrt{|z_{s,0,r}|}&|z_{s,0,r}k^2+g_{2,r}|<1\end{cases}
\ee
containing a dimensionless parameter $c_s$ is chosen to smooth out possible irregularity at the onset of the condensate.

\subsection{Renormalization group flow}
\begin{figure}
\includegraphics[scale=.6]{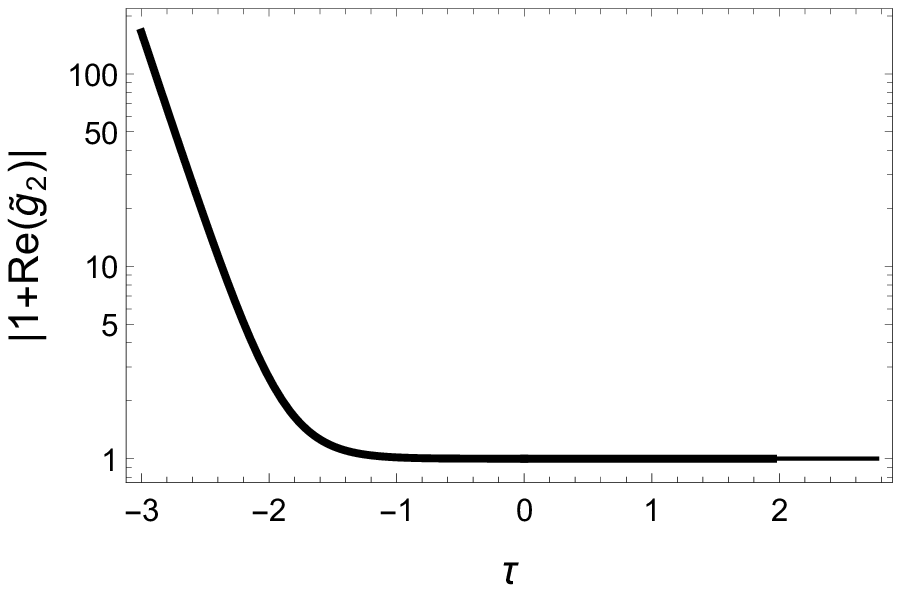}\hskip.5cm\includegraphics[scale=.6]{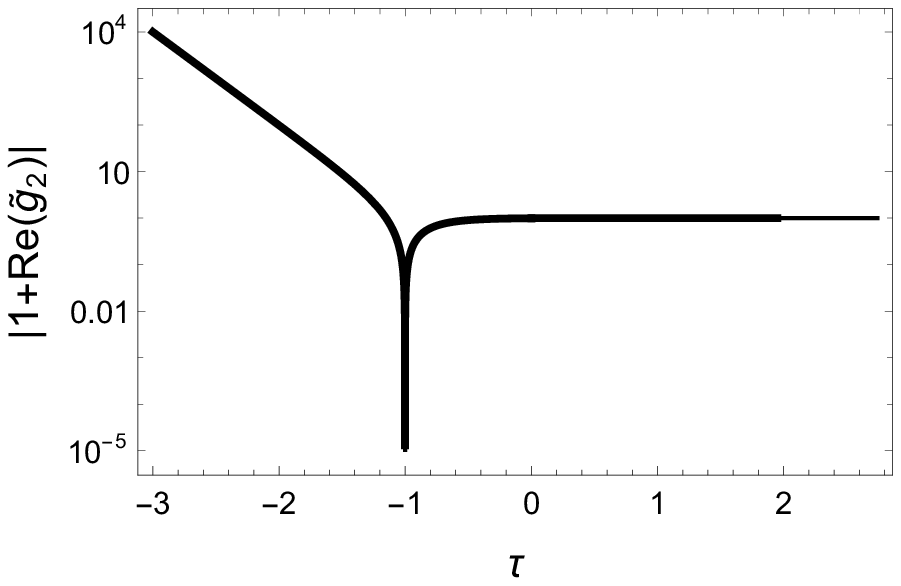}

\includegraphics[scale=.6]{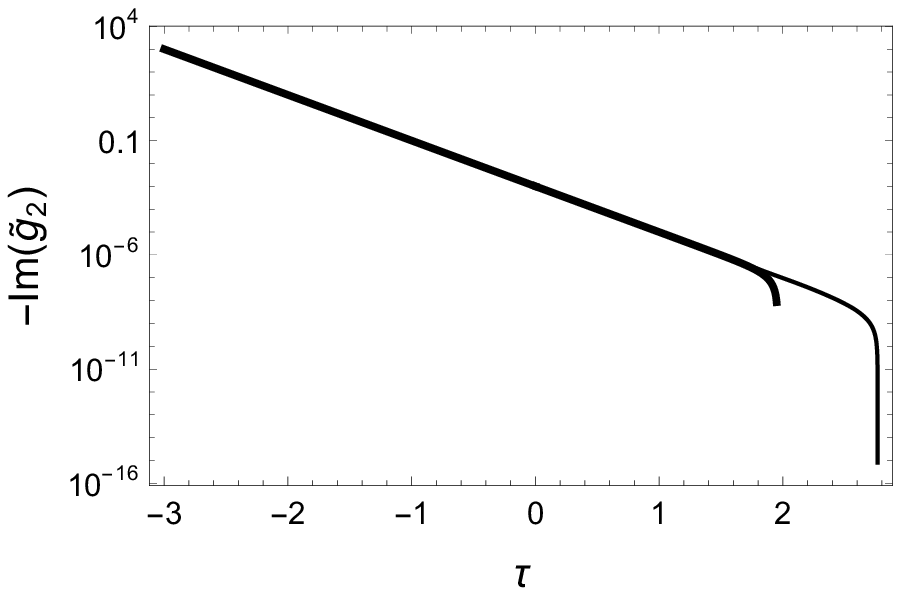}\hskip.5cm\includegraphics[scale=.6]{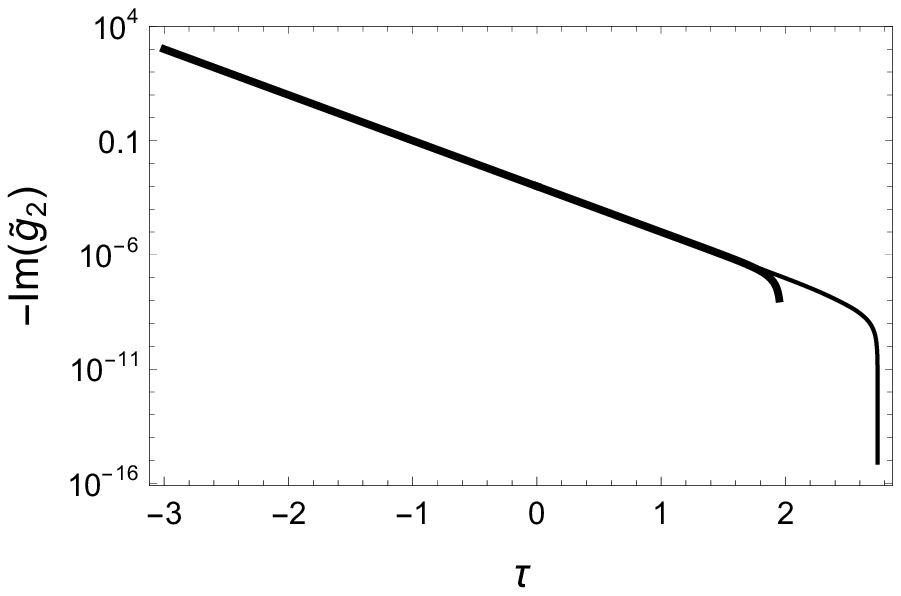}

\includegraphics[scale=.6]{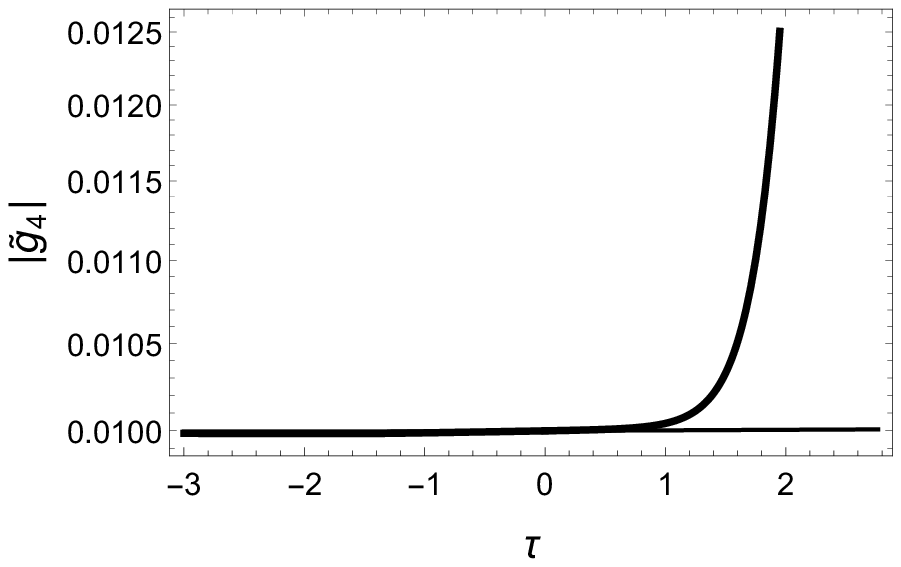}\hskip.5cm\includegraphics[scale=.6]{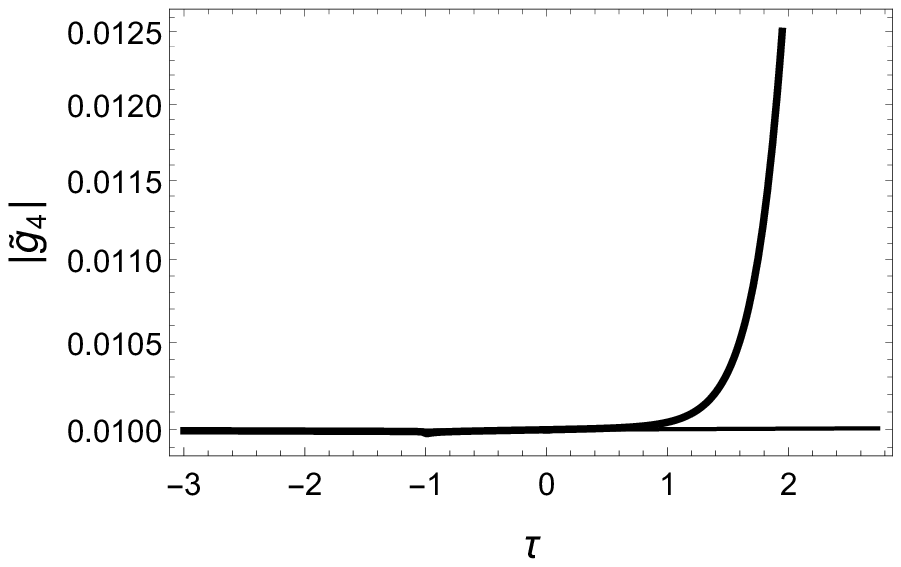}

\includegraphics[scale=.6]{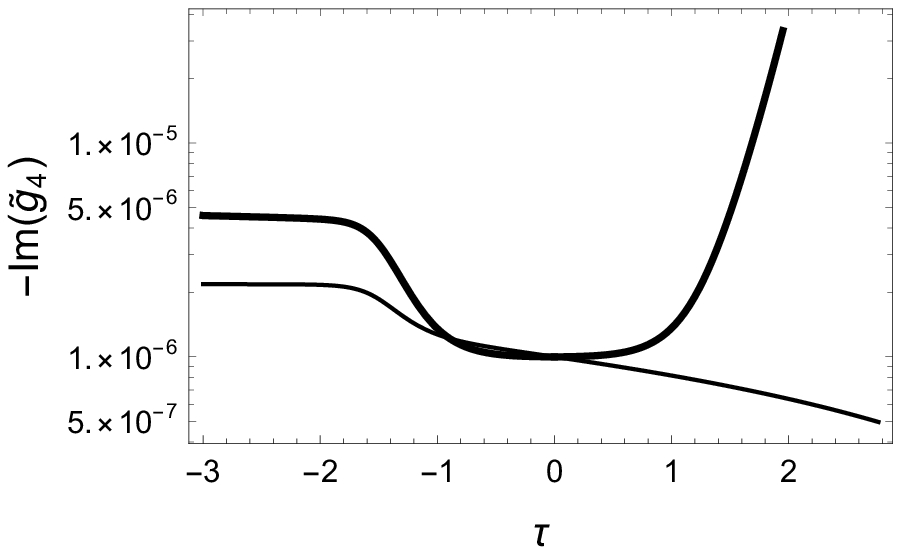}\hskip.5cm\includegraphics[scale=.6]{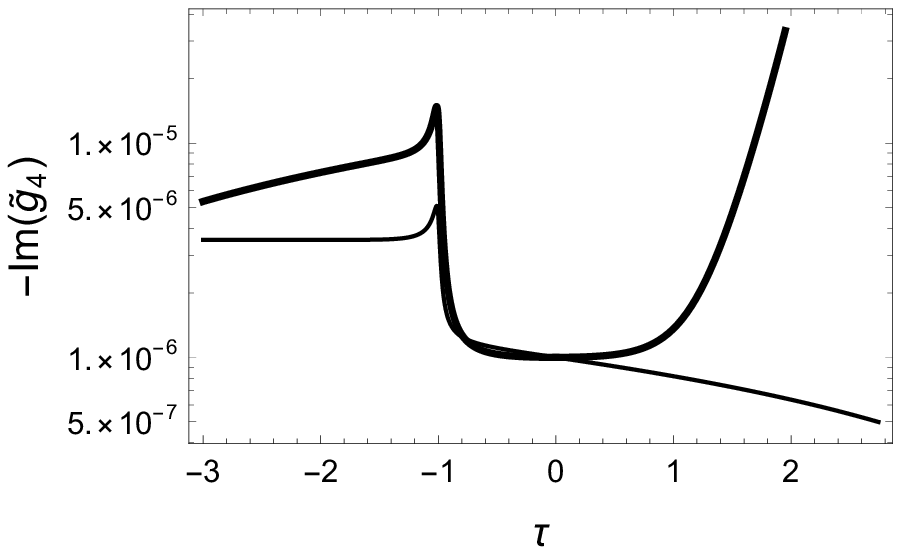}

\hskip.9cm(a)\hskip5.4cm(b)
\caption{The trajectories of the local potential.}\label{uf}
\end{figure}

\begin{figure}
\includegraphics[scale=.6]{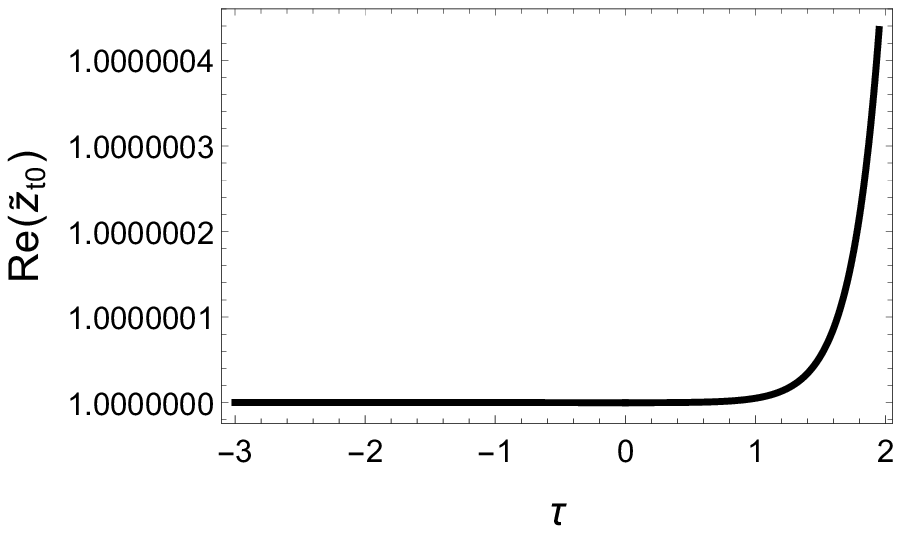}\hskip.5cm\includegraphics[scale=.6]{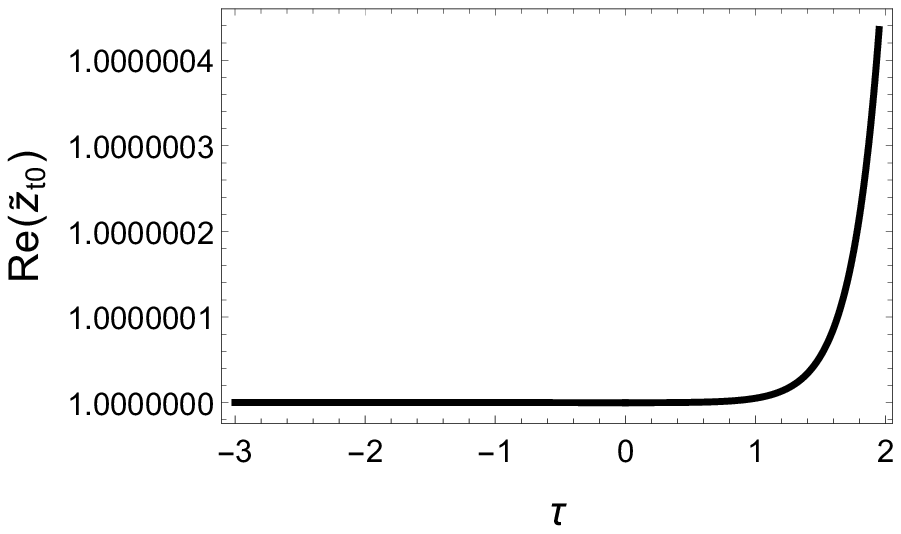}

\includegraphics[scale=.6]{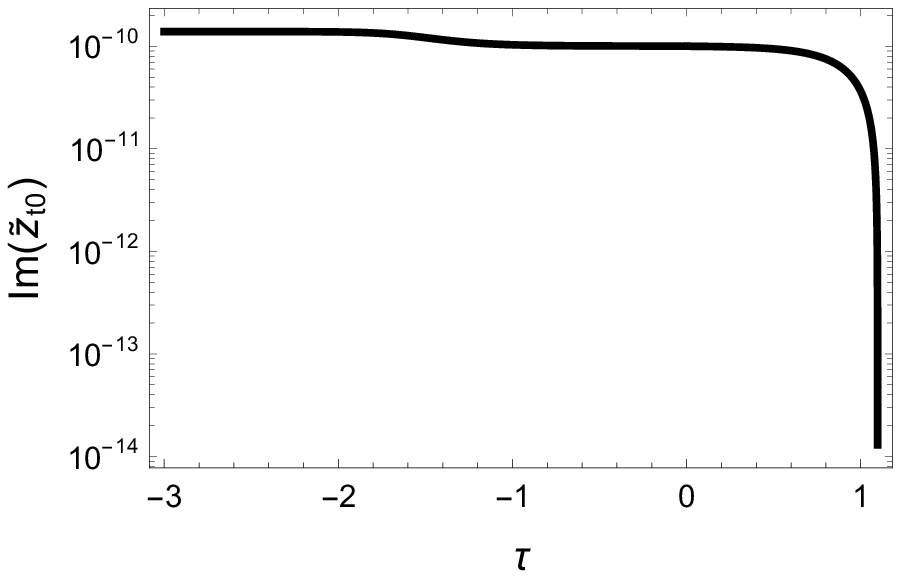}\hskip.5cm\includegraphics[scale=.6]{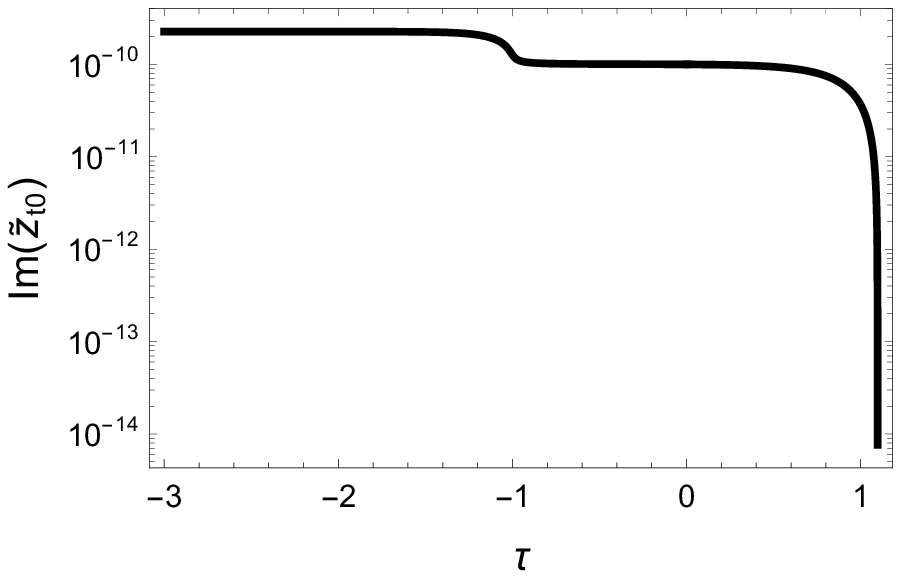}

\includegraphics[scale=.6]{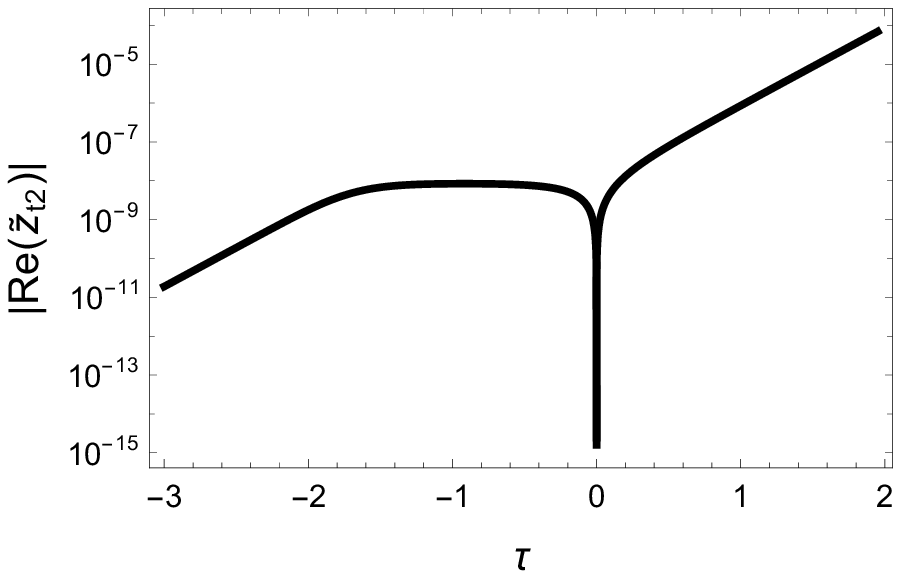}\hskip.5cm\includegraphics[scale=.6]{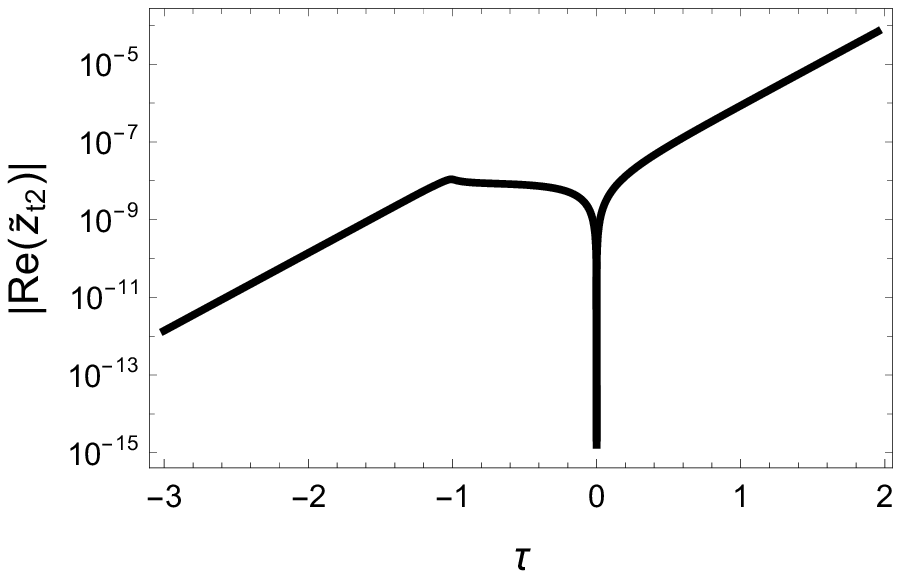}

\includegraphics[scale=.6]{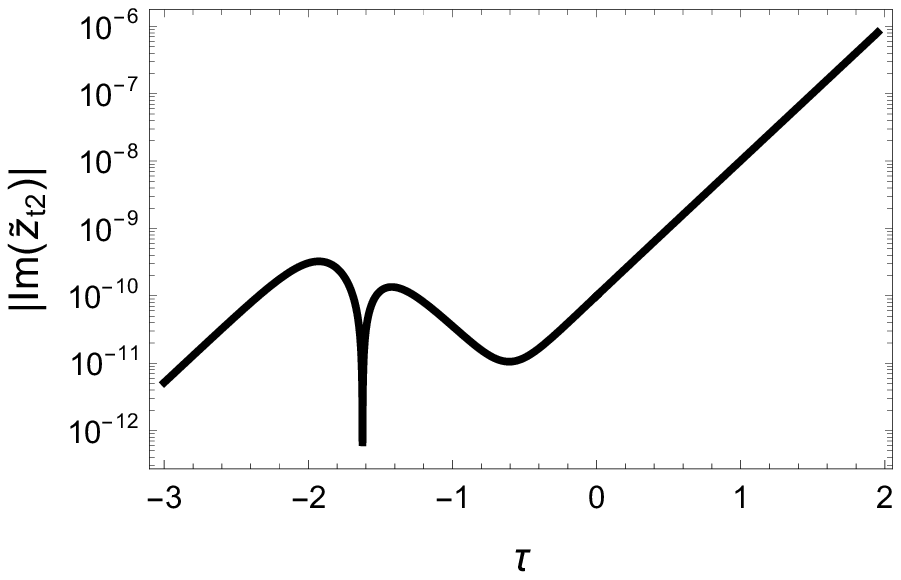}\hskip.5cm\includegraphics[scale=.6]{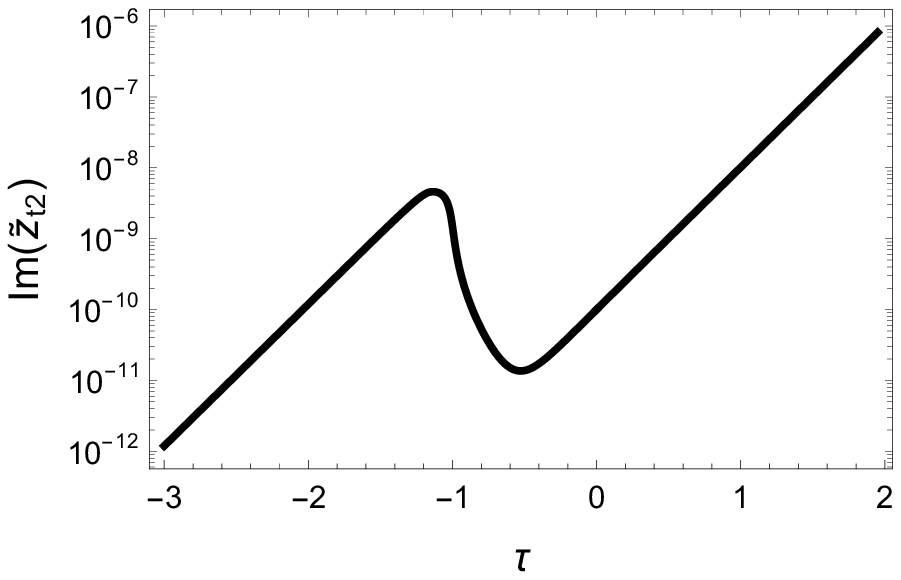}

\hskip.9cm(a)\hskip5.4cm(b)
\caption{The trajectories of wave function renormalization constant.}\label{zf}
\end{figure}

Two typical renormalized trajectories, obtained by integrating numerically the evolution equations from the initial conditions $k=1$ ($\tau=0$) up and downward in the cutoff, are shown below with $\Omega_k=10^4\omega_s$, $c_s=3$ for the dimensionless parameters $\tilde g_2=g_2/k^2$, $\tilde g_4=g_4$, $\tilde Z_t(\tilde\phi)=Z_t(\phi)$, $\tilde\phi=\phi/k$  by the fat line on Figs. \ref{uf}-\ref{zf} for the initial conditions $\tilde Z_t=1+10^{-10}i(1+\tilde\phi^2)$, $\tilde Z_s=1$, $\tilde g_{2,i}=-0.001$, $\tilde g_4,=2-10^{-10}i$ and (a) $\tilde g_{2,r}=0.0001$; (b) $\tilde g_{2,r}=-0.01$. The wave function renormalization constant remains unchanged, $Z_s=1$, when it is field independent in the initial condition. The thin line corresponds to the local potential approximation (LPA) where the running action is truncated to $Z_t=Z_s=1$. The integration of the full evolution equation with wave function renormalization constant was carried out up to $\Lambda_2$, c.f. section \ref{msrss}.

Let us comment first the trajectory of the local potential displayed in Fig. \ref{uf}. The initial conditions at $\tau=0$ are selected according to the realization of the symmetry $\phi\to-\phi$: The theory is in the symmetrical phase for (a), $\tilde g_{2,r}\to\infty$ as $k\to0$ and the trajectory (b) belongs to symmetry broken vacuum with $\tilde g_{2,r}\to-\infty$ as $k\to0$. The IR scaling of $g_{2,r}$ is dominated by the trivial classical contribution, $|g_2|\sim1/k^2$ and $g_{2,i}$ remains relatively independent of the symmetry of the vacuum and is driven by the classical scaling only. The coupling constant shows more structure, $g_{4,i}$ revels that the IR-UV crossover actually consists of two characteristic scales in real time dynamics, and $g_{4,i}$ changes rapidly between them. It is natural to assume that the fast decrease of the life-time of two-particle resonance states form the intermediate scaling regime. While the wave function renormalization constant does not make much difference in the IR direction the UV scaling is strongly modified by the non-renormalizable parameter $z_{t,2}$ since it deflects the trajectory from the Gaussian fixed point, seen clearly on the plots of $|g_4|$. Furthermore it lowers the scale where $\sign(g_{2,i})$ flips and the quasi particles acquire long life-time.

The role of the wave function renormalization constant is easiest to understand by inspecting the trajectory of $z_{t,2}$ because the $\phi$-dependent part drives the evolution of $Z$. Both the real and the imaginary parts of $z_{t,2}$ follow the classical scaling law, $z_{t,2}=\ord{k^2}$, apart of some interruptions which introduce characteristic scales. The initial conditions force $z_{t,2,r}=0$ at $\tau=0$ but $z_{t,2,r}$ regains $z_{t,2,r}\sim-k^2$ as we move to the UV direction after a short shooting up. The continuation towards the IR direction of this short, fast changing behaviour quickly settles to a plateau at a positive value and the classical scaling $z_{t,2,r}\sim k^2$ continues on the positive side as soon as we leave the UV scaling regime. The importance of $g_{t,2,r}<0$ above the initial conditions is that the wave function renormalization constant drives the dynamics towards stronger coupling at large  cutoff. We can not reach the Landau peak with the present simple calculation because the stronger interactions generate condensate by flipping the sign of $g_{2,i}$. The intermediate scaling between the asymptotic UV and IR scaling regimes originates from the non-monotonic feature of $z_{t,2,i}$. There is another turn of the trajectory of the symmetric phase because $z_{t,2,i}$ crosses zero in the intermediate regime and the classical scaling is realized in the asymptotic IR regime on the negative side, $z_{t,2,i}\sim-k^2$. The marginal, renormalizable parameter $z_{t,0}$ follows the classical scaling, $z_{t,0}=\ord{k^0}$ with little variation in the IR but the stronger interaction, induced by $z_{t,2}$ generates more visible scale dependence as we start to move in the UV directions. We found no trajectory which would extend to $k\to\infty$ without flipping the sign of $g_{2,i}$ or $g_{4,i}$. However one should keep in mind that the integration of the evolution equation upward in the cutoff is not about renormalizability.

\section{Summary}\label{concl}
Different aspects of the renormalization group method in Minkowski space-time are touched upon in this work within the four dimensional $\phi^4$ scalar model. The running parameters, defined by the appropriate vertex functions evaluated at the subtraction point, are found complex. The Euclidean theories possess the discrete symmetry under the complex conjugation of the parameters of the action. The symmetry is broken explicitly by the factor $i$ in front of the Minkowski action in the exponent of the path integral but can be recovered by upgrading the complex conjugation to time inversion. Hence the imaginary part of the running parameters indicate the dynamical breakdown of the time reversal symmetry, reflected by the finite life-time of the quasi-particles.

The pole of the resummed propagator is closer to the real than the imaginary energy axis in weakly coupled theories. Thus the choice of the subtraction point is more important than in Euclidean theories. It is natural to choose the subtraction points within a kinematical regime dominated by quasi-particle excitations. In turn this choice renders the Wick rotation singular and requires to work with real time dynamics in the renormalization group method from the very beginning. 

The main novelty of the real time dynamics is that it makes the renormalized trajectory complex. Since the imaginary part of the parameters are generated by the interactions their sign changes easier than for the real part and each sign change introduces a characteristic scale, associated to inhomogeneous saddle points. The path integral is well defined only when the imaginary part of the highest order coupling constant is negative. In our simple calculation the saddle point was assumed to be vanishing and no trajectory was found which can be extended with convergence path integral and without condensate to arbitrary high cutoff.

These results are preliminary, suggesting the need of a more thorough construction of the functional renormalization group method in Minkowski space-time. Few questions, calling for further inquires are the following:

The usual strategy of the renormalization group to obtain Green's functions is the successive elimination of the modes in the path integral. This is certainly a mathematically correct way to deal with multi-dimensional integrals however there is no way to interpret the blocked quantum field theory with lowered cutoff in physical terms. The reason is that the cutoff theory always describes an open dynamics hence its handling requires the Closed Time Path formalism. The quantum fluctuations of the bra and the ket components are independent in a closed dynamics and lead to a formal redoubling of the dynamical variables. These copies of the IR field become coupled by the IR-UV entanglement, leading to involved scaling laws. The extension of the steps, followed in this work, over the Closed Time Path formalism is necessary to handle the open dynamics of cutoff theories, to calculate the expectation values and to find the physically relevant saddle points. A thorough analysis of the phase structure and the possibility of removing the cutoff is needed in this formalism.

The real time dynamics confronts us with an unexpected complication in Quantum Field Theories, namely the difficulties of finding boost invariant non-perturbative regulator \cite{boostinv}. This problem is circumvented here by employing the sharp momentum cutoff, a non-relativistic regulator and the action is projected onto a relativistically invariant form after each blocking. This problem should be clarified rather than swept under the rug and a careful analysis of the status of the boost invariance is required.

\section*{ACKNOWLEDGMENTS}
S. Nagy acknowledges financial support from the Hungarian National Research, Development and Innovation Office NKFIH (Grant No. KH126497).

\end{document}